\documentclass[aps,prd,twocolumn,showpacs]{revtex4}

\usepackage{amssymb}
\usepackage{graphicx}

\newcommand{\md}{{\mathrm{d}}}

\begin{document}

\title{\Large \bf Statistical moments for classical and quantum dynamics: formalism and generalized uncertainty relations}

\author{David Brizuela}
\email{david.brizuela@ehu.es}


\affiliation{Fisika Teorikoa eta Zientziaren Historia Saila, UPV/EHU, 644 P.K., 48080 Bilbao, Spain}
\affiliation{Institut f\"ur Theoretische Physik, Universit\"at zu K\"oln, Z\"ulpicher Stra{\ss}e 77, 50937 K\"oln, Germany}

\begin{abstract}
The classical and quantum evolution of a generic probability distribution is analyzed. To that end,
a formalism based on the decomposition of the distribution in terms of its statistical moments is used,
which makes explicit the differences between the classical and quantum dynamics. In particular, there
are two different sources of quantum effects. Distributional effects, which are also present in the classical
evolution of an extended distribution, are due to the fact that all moments can not be vanishing
because of the Heisenberg uncertainty principle. In addition, the non-commutativity of the basic quantum
operators add some terms to the quantum equations of motion that explicitly depend on the Planck constant
and are not present in the classical setting. These are thus purely-quantum effects.
Some particular Hamiltonians are analyzed that have very special properties regarding the evolution
they generate in the classical and quantum sector.
In addition, a large class of inequalities obeyed by high-order statistical moments,
and in particular uncertainty relations that bound the information that is possible to obtain from a
quantum system, are derived.
\end{abstract}

\maketitle

\section{Introduction}

The classical limit of quantum mechanics is very subtle. One of the main
problems to define such a limit is the difference between the formalisms
used to describe these theories.
Quantum mechanics is analytical since it is based on Hilbert spaces and
operators acting thereon, whereas classical mechanics is geometrical:
it is defined on symplectic manifolds and with Hamiltonian vector fields
describing the evolution. In addition, apart from the purely quantum and purely classical
theories, there are also different attempts to construct hybrid theories that consider
a direct coupling between classical and quantum degrees of freedom, see e.g. \cite{BoTr88, And95, ShSu78, PeTe01, Elz12, CHS12}.
Nevertheless, many of these theories suffer from several drawbacks \cite{CaSa99} and will not be considered here; see
Ref. \cite{BCG12} for a discussion on these hybrid theories within a formalism similar
to the one presented in this paper.

The intuitive idea one usually has in mind about a semiclassical state is that of a
peaked coherent state, whose centroid (the coordinates on the phase space of the expectation
value of the position and momentum operators) follows a classical orbit in phase space. Nonetheless, in Ref. \cite{BYZ94}
a different idea was presented: the classical limit of a quantum system is not a single classical orbit, but an ensemble of orbits.
Therefore, even if it is not peaked and its centroid does not follow a classical trajectory, a quantum state can behave essentially classically if it follows the evolution given by the Liouville equation for an equivalent classical ensemble.
In particular it is well known that generically the centroid of a quantum state
does not follow a classical trajectory. But this very same thing happens with the evolution of the centroid
of a classical distribution; so such an effect can not be regarded as completely quantum.

Another reason to work with classical probability distributions is that, even if in the context
of classical mechanics in principle one could have a point in the phase space as initial condition,
in practice there are always
measurement errors that introduce some uncertainty on the knowledge about the initial
state. Hence, this issue forces us to consider also the evolution of extended probability
distributions on a purely classical setting.

The difference between a quantum and classical distributional evolution of an initial state
could be of particular importance in the context of quantum cosmology. Leaving aside theories that involve multiverses,
there is only one realization of the universe and thus one dynamical state that describes it.
If, at some point, we were able to measure some properties of that state, it will be necessary to compare its classical
and quantum evolution in order to know if the universe is indeed behaving quantum
mechanically or this distributional behavior is just due to the error on our measurements.  

As will be made explicit in this paper, a formalism very well suited to compare the classical
and quantum evolutions of a given physical system is the one developed in \cite{BM98} for
Hamiltonians corresponding to a particle on a potential. In that paper a decomposition of the wave function
into its infinite set of statistical moments is considered. The moments contain the same
physical information as the wave function, but have the great advantage of being observable.
The evolution equations of these variables give rise to the Hamilton equations for
the expectation values of the position and momentum operators, with corrections coming from
the moment variables. On the other hand, in the same way as for the quantum theory,
it is possible to define the moments corresponding to a classical ensemble and obtain their evolution equations.
In this way, the classical and quantum evolution of a given physical system is described
in a very similar setting and the comparison between them is straightforward.

A similar formalism to the one developed in \cite{BM98},
but with a different ordering of the basic variables, was presented in \cite{BoSk06} for generic Hamiltonians
in a canonical framework.
This formalism has been widely applied to different cosmological models (see \cite{Boj12} for a recent review).
One of the advantages of this approach is that it can be applied also
when the dynamics is described by a Hamiltonian constraint, as opposed to a Hamiltonian function \cite{BoTs09}.
In this context, isotropic cosmological models with negative and positive cosmological constant have been
studied in Ref. \cite{BoTa08} and \cite{BBH11} respectively. It has also been applied to simple
bounce scenarios in loop quantum cosmology \cite{Boj07}. Furthermore,
the problem of time in semiclassical regimes, as well as the relational quantum dynamics, has also been considered
in \cite{BHT11, HKT12}.

In the present paper the classical counterpart to the quantum formalism developed in \cite{BoSk06}
will be presented. In this way, a formalism similar to the one introduced in \cite{BM98} will
be obtained, but valid for generic Hamiltonians and with a different ordering of the basic variables. In addition,
due to the generic formula between two moments obtained in \cite{BSS09, BBH11}, much more
compact  evolution equations will be presented, which allow for an easier understanding of
the terms involved. Harmonic and linear Hamiltonians will be considered due to their special
properties regarding the classical and quantum evolution they generate. Furthermore, making
use of the Cauchy-Schwarz inequality, high-order inequalities obeyed by statistical moments
will be systematically obtained. In the quantum case, a specific subset of these inequalities
will give rise to high-order uncertainty relations, that generalize the well-known Heisenberg
uncertainty principle.

The rest of the article is organized as follows. In Sec. \ref{sec_formalism} the general
formalism is explained. In Subsec. \ref{sec_formalism_quantum} a brief summary of the
quantum formalism introduced in \cite{BoSk06} is given, whereas in
Subsec. \ref{sec_formalism_classical} its classical analog is developed. The
rest of this section underlines the differences between the classical and quantum settings,
and analyzes how to obtain dynamical and stationary states in this context.
Section \ref{sec_specialhamiltonians} discusses the special classical and quantum
behavior of harmonic and linear Hamiltonians. In Sec. \ref{sec_inequalities}
high-order inequalities obeyed by both classical and quantum statistical moments
are obtained and analyzed. Finally, Sec. \ref{sec_conclusions} presents the conclusions
and summarizes the main results of the article.

\section{General formalism}\label{sec_formalism}

\subsection{Quantum moments}\label{sec_formalism_quantum}

In this subsection the formalism developed in \cite{BoSk06} is briefly reviewed.
Let us assume a quantum mechanical system with one degree
of freedom described by the basic conjugate operators $(\hat q, \hat p)$. We define the quantum moments
\begin{equation}
G^{a,b}:=\langle(\hat p - p)^a \,(\hat q - q)^b\rangle_{\rm Weyl},
\end{equation}
where $p:=\langle\hat p\rangle$, $q:=\langle\hat q\rangle$ are the
expectation values of the momentum and position operator, respectively,
and the subscript Weyl stands for totally symmetric
ordering. The sum between its two indices $(a+b)$ will be referred as
the order the moment $G^{a,b}$ .

Note that through this decomposition the wave function $\Psi(q,t)$ gets
replaced
by its infinite set of statistical moments $G^{a,b}(t)$, which only depend on time.
This is quite similar to what is done in physical
problems with certain symmetry: use special functions
adapted to that symmetry and remove the dependence on the
trivial directions, like for example spherical harmonics
when dealing with spherical symmetry.

Performing the Taylor expansion of the Hamiltonian operator $\hat H$, an effective
Hamiltonian $H_Q$ is obtained as function of the expectation values and moments:
\begin{eqnarray} \label{HQ}
H_Q(q,p,G^{a,b})&=&\langle\hat{H}(\hat q, \hat p)\rangle_{\rm Weyl}
\nonumber\\\nonumber
&=&\langle\hat{H}(\hat q-q+q, \hat p -p +p)\rangle_{\rm Weyl}\\\nonumber
&=& \sum_{a=0}^\infty\sum_{b=0}^\infty \frac{1}{a!b!}
\frac{\partial^{a+b} H}{\partial p^a\partial q^b} G^{a,b}\\
&=&H(q,p) + \sum_{a+b\geq 2}\frac{1}{a!b!}\frac{\partial^{a+b} H}{\partial p^a\partial q^b} G^{a,b}.
\end{eqnarray}
The classical Hamiltonian $H(q,p)$ is obtained by replacing $\hat q$ and $\hat p$ by their corresponding expectation values
$q$ and $p$, respectively, in the explicit expression of the Hamiltonian operator $\hat H(\hat q, \hat p)$.
It can be shown that, if one defines the Poisson bracket for expectation values of
arbitrary operators $\hat f$ and $\hat g$ by
the relation $\{\langle\hat f\rangle,\langle\hat g\rangle\}=
-i\hbar^{-1}\langle[\hat f,\hat g]\rangle$,
the evolution generated by this Hamiltonian is equivalent to
the Schr\"odinger flow of quantum states. Thus, this effective Hamiltonian
encodes the complete dynamical information of our variables: the expectation values $(q,p)$ and the infinite set of moments.
With the mentioned definition, it is immediate to see that, given the commutation relation $[\hat q,\hat p]=i\hbar$, the Poisson
bracket between conjugate variables reduces to the canonical one: $\{q,p\}=1$.
Moreover, one can easily show that moments have vanishing Poisson brackets with the
basic expectation values:
\begin{equation}
 \{G^{a,b},p\}=0=\{G^{a,b},q\}\,.
\end{equation}

The general Poisson brackets between moments is more involved, but a general
formula is given by,
\begin{eqnarray}\label{GGbrackets}
&&\{G^{a,b},G^{c,d}\}=
a\, d \,G^{a - 1, b} \,G^{c, d - 1} - b \,c\, G^{a, b - 1}\, G^{c - 1, d}
\nonumber\\
&&+\sum_{m=0}^{\lfloor\frac{M-1}{2}\rfloor}
(-1)^m\frac{\hbar^{2m}}{2^{2m}}
K_{abcd}^{2m+1}\, G^{a+c-2m-1, b+d -2m-1},
\end{eqnarray}
where $M:={\rm Min}(a + c, b + d, a + b, c + d)$ and
the coefficients
\begin{eqnarray}
K_{abcd}^n =\!\!
\sum_{m= 0}^{n}
(-1)^m
m!(n-m)!
\!\left(\!\!\begin{array}{c}
a\\m
\end{array}\!\!\right)\!
\left(\!\!\!\begin{array}{c}
b\\n-m
\end{array}\!\!\!\right)\!
\nonumber
\left(\!\!\!\begin{array}{c}
c\\n-m
\end{array}\!\!\right)\!
\left(\!\!\!\begin{array}{c}
d\\m
\end{array}\!\!\right)
\end{eqnarray}
have been defined. Note that this expression mixes different orders
in a nontrivial way since the bracket between two moments of order $(a+b)$ and $(c+d)$,
respectively, are a combination of moments from at most order $|(a+b)-(c+d)|$ (or higher,
depending which of the combinations in the definition of $M$ is the minimum),
up to order $(a+b+c+d-2)$.

The evolution equations for different variables are then obtained by simply
computing their Poisson brackets with the effective Hamiltonian (\ref{HQ}).
In this way, it is straightforward to get the following relations for
the expectation values:
\begin{eqnarray}\label{eqqq}
\frac{dq}{dt}&=&\{q,H_{ Q}\}\nonumber\\&=&\frac{\partial H(q,p)}{\partial p}+\sum_{a+b\geq2}\frac{1}{a!b!}\frac{\partial^{a+b+1}H(q,p)}{\partial p^{a+1}\partial q^{b}}G^{a,b},\\
\frac{dp}{dt}&=&\{p,H_{ Q}\}\nonumber\\&=&-\frac{\partial H(q,p)}{\partial q}-\sum_{a+b\geq2}\frac{1}{a!b!}\frac{\partial^{a+b+1}H(q,p)}{\partial p^{a}\partial q^{b+1}}G^{a,b}.
\label{eqpq}
\end{eqnarray}
Note that if the classical Hamiltonian $H(q,p)$ is at most quadratic in the variables, the sums
in these equations will give no contribution and the expectation values will
exactly fulfill their corresponding classical equations of motion. In Sec. \ref{sec_harmonic} this
and other special features of quadratic Hamiltonians will be analyzed.

On the other hand, the equations of motion for the moments can be written as follows:
\begin{eqnarray}\label{generaleqG}
\frac{d G^{a,b}}{dt}=\{G^{a,b},H_Q\}
=\sum_{c+d\geq 2} \frac{1}{c!d!}
\frac{\partial^{c+d} H}{\partial p^c\partial q^d} \{G^{a,b},G^{c,d}\},
\end{eqnarray}
where the Poisson bracket (\ref{GGbrackets}) should be replaced.
As can be seen, in principle (except for certain particular forms of
the Hamiltonian $H$) this equation will get contributions from several
orders. 
 
Even if the infinite set of equations is equivalent to the Schr\"odinger equation,
for practical reasons in general, in order to analyze the dynamics of the system,
it will be necessary to introduce a cut-off $>N$. That is, a maximum order $N$,
so that all $G^{a,b}$ are assumed to be vanishing if $a+b>N$.
Due to the special form of
the brackets (\ref{GGbrackets}), that mix higher- and lower-order
moments, the introduction of such a
cut-off could be made in two different and inequivalent ways regarding
the equations of motion for the moments (\ref{generaleqG}).
1/ Truncate the Hamiltonian at order $N$,
calculate the equations of motion (\ref{generaleqG}), and truncate
again the right-hand side of these equations. 2/ Consider the complete
Hamiltonian $H_Q$ without truncation, calculate the equations
of motion (\ref{generaleqG}) and truncate then the result at order $N$.
(In practice, for this second procedure it is enough to consider the
Hamiltonian up to order $2 N$ since higher orders of the Hamiltonian
will only introduce moments of an order greater than $N$ in the equations,
which will be made to vanish when truncating the result.) On the other hand,
the equations of motion for the expectation values $q$ and $p$ do not
depend on the way the truncation is done. The truncation of the Hamiltonian
at the desired order $N$ is sufficient to get only moments up to that
order in their equations of motion and considering the full Hamiltonian
$H_Q$ will not introduce new terms.

In principle, from a perturbative perspective, one would say that the
first route is more consistent: no knowledge of higher-order terms
is made use of at any step. Nevertheless, this is a very peculiar
perturbative expansion since, as commented above, the Poisson
brackets of a moment of order $N$ with a moment of order $N+1$
does not generate only moments of order $N$ and greater;
lower-order terms also appear. Following the second route guarantees
that the equations of motion truncated at a given order $N$ will
contain all contributions from moments up to that order. That is,
when considering the truncation at order $N+1$, the lower-order
terms will not change.

This formalism is very practical because one deals directly with 
measurable quantities (expectation values) instead of with the wave function
and it is immediate to see the corrections for classical equations of motion.
These corrections are usually regarded as purely quantum but note
that there is no explicit $\hbar$ involved in the definition of the moments or
the expansion of the Hamiltonian. Only the Poisson brackets between
two moments introduces $\hbar$ terms. A question then arises: is it
possible to obtain a similar formalism for classical mechanics? As
will be explained in the next section, the answer is in the affirmative.

\subsection{Classical moments}\label{sec_formalism_classical}

As explained in \cite{BM98},
it is possible to define the classical analog of the moments define in the previous subsection.
Let us assume a classical
mechanical system described by the Hamiltonian $H$ with a phase
space coordinatized by the conjugate pair $\tilde q$ and $\tilde p$ with Poisson
bracket $\{\tilde q,\tilde p\}_c=1$ (the subscript $c$ is used not to confuse this bracket
with the one defined previously for expectation values). A classical ensemble is defined via a probability distribution
$\rho(\tilde q,\tilde p,t)$ on the classical phase space. The evolution of this distribution is
given by the Liouville equation,
\begin{equation}\label{liouville}
\frac{\partial \rho}{\partial t}=-\{\rho,H(\tilde q,\tilde p)\}_c,
\end{equation}
which asserts that the probability distribution is conserved through physical
trajectories $d\rho/dt=0$.

In order to explode further the parallelism between the classical and quantum
worlds, the
classical expectation value for any function in the phase space $f(\tilde q,\tilde p)$
can be defined as
\begin{equation}
\langle f(\tilde q,\tilde p) \rangle_{\rm c} := \int d\tilde q d\tilde p f(\tilde q,\tilde p) \rho(\tilde q,\tilde p,t),
\end{equation}
where the integration extends to the whole domain of the probability distribution on the phase space.

Making use of this operation, the classical mean values for
the position $\tilde q$ and momentum $\tilde p$ are defined:
\begin{eqnarray*}
q(t):=\langle \tilde q \rangle_{\rm c},\\
p(t):=\langle \tilde p \rangle_{\rm c}.
\end{eqnarray*}
These are the coordinates on the phase space of the centroid of our distribution.
The classical moments are further defined as
\begin{equation}
C^{a,b}:=\langle (\tilde p-p)^a (\tilde q-q)^b \rangle_{\rm c}.
\end{equation}
Since all objects in this expression commute, contrary to the quantum case, the ordering is absolutely
irrelevant. Note that the expectation value of any function can be regarded
as a function of the coordinates of the centroid $(\tilde q,\tilde p)$ and the moments $C^{a,b}$.
To make this explicit it is enough to make a Taylor expansion around the position
of the centroid:
\begin{eqnarray}
\langle f(\tilde q,\tilde p) \rangle_{\rm c} &=& \langle f(q+\tilde q-q,p+\tilde p-p) \rangle_{\rm c}\nonumber\\&=&
f(q,p) + \sum_{a+b\geq 2}\frac{1}{a!b!}\frac{\partial^{a+b} f(q,p)}{\partial q^a\partial p^b}C^{a,b}.\,\,\,
\end{eqnarray}

In order to obtain the equations of motion for different variables,
one should define the following Poisson bracket between expectation values,
\begin{equation}\label{defbracket}
\{ \langle f(\tilde q,\tilde p)\rangle_c,\langle g(\tilde q,\tilde p) \rangle_c \}:=
 \langle \{ f(\tilde q,\tilde p), g(\tilde q,\tilde p) \}_c\rangle_c.
\end{equation}
This operation has the good properties one expects for a bracket and gives
the expected results for the expectation values of fundamental variables,
now defined as the coordinates of the centroid of the distribution:
\begin{eqnarray}
\{q,p\}=\langle\{\tilde q,\tilde p\}_c\rangle_c=1,\\
\{q,q\}=0=\{p,p\}.
\end{eqnarray}
As in the quantum case, the classical moments commute with these mean values,
\begin{equation}
\{ C^{a,b},q \}=0=\{ C^{a,b},p \},
\end{equation}
and the Poisson bracket between any two classical moments is given
as follows:
\begin{eqnarray}
\{C^{a,b},C^{c,d}\}&=&
a\, d \,C^{a - 1, b} \,C^{c, d - 1} - b \,c\, C^{a, b - 1}\, C^{c - 1, d}
\nonumber\\\label{CCbrackets}
&+&\, (b\,c-a\,d)\, C^{a+c-1, b+d-1}.
\end{eqnarray}
Comparing these equations with the corresponding to the quantum moments (\ref{GGbrackets}),
it is possible to see that the key difference between classical and quantum moments
lies in the non-commutativity of the quantum operators. The first two summands of Eqs.
(\ref{GGbrackets}) and (\ref{CCbrackets}) are identical, but the last term differs. In the quantum
case there is a sum in even powers of $\hbar$ whereas in the classical case only the first
summand, which is independent of $\hbar$, is kept. In the quantum case, these terms appear
due to the non-commutativity of the basic operators and can be different for different orderings chosen in the
definition of the moments. Nevertheless, it is not possible to make these terms vanishing
by a redefinition of the ordering \cite{BM98}. As a side remark, note that the
Poisson bracket between a second-order moment and any other moment give the same result
in the classical or in the quantum case. As it will be explained below, this fact will make
the quantum evolution of the system up to second order completely equivalent to the classical evolution.

With these definitions at hand, the evolution of any expectation value $\langle f\rangle_c$
can be obtained in the following way,
\begin{eqnarray}
\frac{d \langle f\rangle_c}{dt}=\frac{d}{dt}\int d\tilde q \,\,d\tilde p \,\,\rho(\tilde q,\tilde p,t)\,\, f(\tilde q,\tilde p)\\
=-\int d\tilde q\,\, d\tilde p\,\,\{\rho(\tilde q,\tilde p,t),H(\tilde q,\tilde p)\}_c\,\, f(\tilde q,\tilde p),
\end{eqnarray}
where use of the Liouville equation has been made. Writing the Poisson bracket as
derivatives with respect to the fundamental variables $(\tilde q,\tilde p)$ and integrating by parts,
it is easy to obtain
\begin{eqnarray}
\!\frac{d \langle f\rangle_c}{dt}\!=\!\!\int\!\! d\tilde q \,d\tilde p \,\rho\,\{f,H\}_c\!=\!
\langle \{f,H\}_c\rangle_c
\!=\!\{\langle f\rangle_c, \langle H\rangle_c\}.
\end{eqnarray}
This shows explicitly that the evolution of the classical ensemble described by the distribution $\rho(\tilde q,\tilde p,t)$
will be given by the expectation value of the classical Hamiltonian, which can be Taylor
expanded around the centroid of the distribution:
\begin{eqnarray}
H_{\rm C}(q,p,C^{a,b})&:=&\langle  H(\tilde q,\tilde p) \rangle_{\rm c}\\
&=&\langle H(q+\tilde q-q,p+\tilde p-p) \rangle_{\rm c}\nonumber\\\nonumber
&=&H(q,p) + \sum_{a+b\geq2}\frac{1}{a!b!}\frac{\partial^{a+b} H(q,p)}{\partial p^a\partial q^b}C^{a,b}.
\end{eqnarray}

In particular, the equations of motion of the centroid are given by
\begin{eqnarray}\label{eqqc}
\frac{dq}{dt}&=&\{q,H_{\rm class}\}\nonumber\\
&=&\frac{\partial H(q,p)}{\partial p}
+\sum_{a+b\geq 2}\frac{1}{a!b!}\frac{\partial^{a+b+1}H(q,p)}{\partial p^{a+1}\partial q^{b}}C^{a,b},\\
\frac{dp}{dt}&=&\{p,H_{\rm class}\}\nonumber\\
&=&\!\!-\frac{\partial H(q,p)}{\partial q}
-\!\!\sum_{a+b\geq 2}\!\frac{1}{a!b!}\frac{\partial^{a+b+1}H(q,p)}{\partial p^{a}\partial q^{b+1}}C^{a,b}.
\label{eqpc}
\end{eqnarray}
It is clear from here that the centroid of the distribution does not generically follow a classical trajectory.
Although in the particular case that one chooses a Dirac delta centered at $(\tilde q(0),\tilde p(0))$
as the initial distribution $\rho(\tilde q,\tilde p,t=0)$ all moments will vanish and this
point would follow a classical trajectory.
The first terms of the right-hand side of both equations are the ones that corresponds to the usual
Hamilton equations, but then there are corrections due to the moments. As in the quantum case,
the moments and the mean values of the distribution form an infinite system of couple differential
equations. The evolution of the classical moments is given by
\begin{equation}\label{eqCc}
\frac{d C^{a,b}}{d t} = \sum_{c+d\geq 2}\frac{\partial^{c+d}H}{\partial p^c\partial q^d}\{C^{a,b},C^{c,d}\}.
\end{equation}
These equations of motion can be recovered from the evolution equations of the quantum moments (\ref{generaleqG})
simply by imposing $\hbar=0$.

\subsection{Classical versus quantum dynamics}

Note that remarkably, due to the properties of the Poisson brackets between quantum expectation values,
there is no $\hbar$ term in the equations for second-order quantum moments (\ref{generaleqG}) nor in the equations
for the expectation values $q$ and $p$ (\ref{eqqq}--\ref{eqpq}).
Therefore, a truncation at second order give the same set of equations both for quantum and
classical variables and thus it is necessary to study at least a third-order
truncation in order to find differences between the classical and quantum evolutions.

The evolution equations for moments of higher order will in general be different due to the
explicit $\hbar$ terms that appear in the equations for quantum moments (\ref{generaleqG}). In addition, all
equations are coupled and when considering a third (or higher) order truncation the evolution
of expectation values ($q$, $p$) and of second-order moments will be affected due to the back-reaction of other variables. In this case their evolution with the classical or quantum system
will differ even if their equations of motion still have the same formal form.

One important conclusion of this analysis is that quantum effects have two different origins.
On the one hand, there are \emph{distributional effects} due to the fact that in the quantum
theory one necessarily deals
with the evolution of a spread probability distribution, as opposed to a point trajectory,
on the phase space and thus moments are non-vanishing. As has been explained, these effects are also present in the classical
setting when considering the evolution of a probability distribution on the phase space as
correction terms to the usual Hamilton equations.
On the other hand, the \emph{non-commutativity} (or \emph{purely-quantum}) \emph{effects} appear in the quantum equations of
motion through explicit $\hbar$ terms. Their origin is the non-commutativity of basic operators
$\hat q$ and $\hat p$, and they are not present in the classical setting. On the contrary
to the distributional effects, that may appear for instance when the initial conditions
of a classical system are known only with certain finite precision, these latter effects
are thus genuinely quantum.

It is difficult to measure the relative strength and importance of each effect since all equations
are coupled and therefore both effects mix together. Intuitively the distributional effects
depend on the peakedness of the distribution since, in general, the more spread the distribution the larger the
magnitude of different moments. The non-commutativity effects enter in the equations of motion as
a power series in $\hbar^2$, so the value of Planck constant can give an estimate of their relative importance.

An idea that will be pursued in a forthcoming article \cite{forthcoming}
is to analyze three different evolutions of the system: the classical and quantum
evolution of the same initial distribution, as well as the classical point trajectory
(taking all moments to vanish). The comparison between the classical point trajectory
and the classical evolution of the expectation values with non-vanishing moments will estimate the strength
of distributional effects, whereas the comparison between the quantum and classical
evolutions will estimate the magnitude of the non-commutativity effects.

\subsection{Stationary and dynamical states}

Once the equations of motion for the Hamiltonian system have been obtained
[Eqs. (\ref{eqqq}-\ref{generaleqG}) for the quantum setting
and (\ref{eqqc}-\ref{eqCc}) for the classical one],
the complete evolution of a dynamical state can be computed by solving them.
The only additional physical input that is needed are the initial conditions
for both the expectation values and the moments.

Nevertheless, the stationary states of a quantum system are also of
remarkable importance. These states are solutions of the time-independent
Schr\"odinger equation and, hence, eigenstates of the Hamiltonian operator.
Their unique time dependence is encoded on a pure-phase
term $e^{-iEt/\hbar}$. In the present formalism, based on a decomposition in moments
of the wave function, it is possible to obtain the stationary
states by analyzing the dynamical system given by the infinite set of
moments (or, approximately, by its truncated version). In fact, there is no
need to solve any differential equation: it is enough to equal to zero
the right-hand side of all equations of motion (\ref{eqqq}-\ref{generaleqG}) and solve the algebraic
system in order to obtain the critical or equilibrium points.
[The same analysis can be applied to Eqs. (\ref{eqqc}-\ref{eqCc}) to obtain the stationary classical distributions.]
These equilibrium points are the equivalent of the quantum stationary states.
In general, these equilibrium points would differ from the ones obtained in the
case of a Dirac delta distribution with vanishing moments $C^{a,b}=0=G^{a,b}$, since the equations
of motion of both $q$ and $p$ get corrections by moments.

However, in some cases the commented algebraic system of equations will not be complete \cite{forthcoming},
in such a way that will not be possible to find all moments,
and additional relations must be found.
Since this stationary states are eigenstates of the Hamiltonian operator,
they obey that $\langle \hat H\rangle = E$, and in fact it is possible
to find easily additional restrictions for them. For instance, the
expectation value $\langle[\hat H,[\hat H, g(\hat q)]] \rangle$ must vanish 
for such a state. For the particular case of  the Hamiltonian of a particle
in a potential, $\hat H= \hat p^2/2+V(\hat q)$, that condition can be rewritten
in the following way \cite{Ban77, RiBl79, BrSu81}:
\begin{equation}
2 E \langle g'' \rangle-2\langle g'' V \rangle -\langle g' V'\rangle
+\frac{\hbar^2}{4} \langle g'''' \rangle=0,
\end{equation}
for any $g(\hat q)$. If the potential is polynomial in $\hat q$, for instance
$V(\hat q)=\hat q^m$, the free function can be chosen as
$g(\hat q)=(\hat q-q)^{n+2}/(n+2)$ and, taking into account the relation
\begin{equation}
 \langle\hat q^n \rangle = \sum_{m=0}^nq^mG^{0,n-m},
\end{equation}
write down a recursive relation for moments of the form $G^{0,n}$. In
the particular case that the expectation value is vanishing $q=0$,
the recursive relation takes the following form:
\begin{eqnarray}\label{recursive}
(2 n+m+2) G^{0,n+m}&=&2 E (n+1) G^{0,n}+\nonumber\\&+&\!\!\frac{\hbar^2}{4}(n+1)n(n-1)G^{0,n-2}.\,\,\,\,\,
\end{eqnarray}

This means that if moments up to order $G^{0,m}$ are known, the
higher-order fluctuations can be obtained through this recursive relation.
The equivalent computation for the classical moments gives the same
recursive relation but with a vanishing $\hbar=0$. Therefore, the classical
moments will obey a simpler (two-point) recursive relation.

Finally, other states of interest are the
coherent states, which are defined as quantum dynamical states with an almost
classical behavior. More precisely, their expectation value follow a
classical trajectory on the phase space, whereas all their quantum moments
are constant. Usually the value of the moments is taken to be of minimal
uncertainty. This kind of states can only be constructed for the harmonic
oscillator exactly. Although it might be possible to introduce a generalization
of these states allowing for a relaxation of some of the commented features.

\section{Special Hamiltonians}\label{sec_specialhamiltonians}

In this section two different class of Hamiltonians will be analyzed
that have a very particular properties regarding the classical and quantum
evolution they generate. On the one hand, as will be explained below, harmonic Hamiltonians (that are
at most quadratic in both basic variables $q$ and $p$) essentially generate
the same classical and quantum dynamics \cite{Has78, ReHe80, AnHa85}. On the other hand, Hamiltonians
that are linear in one of the basic variables induce exactly the same
classical and quantum evolution on the expectation values $q$ and $p$,
as well as on an infinite set of moments.

\subsection{Harmonic Hamiltonians}\label{sec_harmonic}

Let us assume that we have a quantum Hamiltonian $H(q,p)$ that is at most quadratic
in both basic variables $q$ and $p$. The effective Hamiltonian will then be given by
\begin{eqnarray*}
H_Q= H(q,p) +\frac{1}{2}\frac{\partial^2 H}{\partial p^2} G^{2,0}
+\frac{\partial^2 H}{\partial q\partial p} G^{1,1}
+\frac{1}{2}\frac{\partial^2 H}{\partial q^2} G^{0,2}.
\end{eqnarray*}

In this particular case the equations of motion both for the classical and quantum
expectation values take the following form:
\begin{eqnarray}\label{harmoniceqq}
\frac{dq}{dt}=\frac{\partial H(q,p)}{\partial p},\\
\frac{dp}{dt}=-\frac{\partial H(q,p)}{\partial q}.
\label{harmoniceqp}
\end{eqnarray}
These are the classical equations of motion for a point on the phase space,
and do not get any correction term from the moments. In addition, the
evolution equation for the moments can be written as
\begin{eqnarray}\label{eq_harmonic}
\frac{d G^{a,b}}{dt}
&=& b \frac{\partial^{2} H}{\partial p^2} G^{a+1,b-1}
+(b-a) \frac{\partial^{2} H}{\partial p\partial q} G^{a,b}\nonumber\\
&-& a \frac{\partial^{2} H}{\partial q^2} G^{a-1,b+1}.
\end{eqnarray}
It is easy to check that
the equations of motion corresponding to the moments
of a classical ensemble have the same structure as those for
quantum variables just by replacing $G^{a,b}$ by $C^{a,b}$ in the last relation.

Therefore the dynamics generated by the harmonic Hamiltonian obey several special properties:
\begin{itemize}
\item On the one hand, the system of equations turns out to be linear
(both in moments as well as in $q$ and $p$ variables) and with constant
coefficients.
Hence, the system can be easily solved analytically.
\item On the other hand, all orders decouple: the equation of motion for a moment
of ${\cal O}(a+b)$ only contains moments of that order.
\item In particular, as we have
already commented, there
appears no moment on the Eqs. (\ref{harmoniceqq}) and (\ref{harmoniceqp}) so there is no back-reaction of
the moments on the trajectories followed by the mean values. Therefore, the
centroid of the distribution will exactly follow a classical dynamical (point)
trajectory in phase space.
\item In the general case, only the full $H_Q$ (and for the classical treatment
$H_C$) is a constant of motion. But in this particular case, due to the decoupling
of the equations of the centroid with the moments, $E_{centroid}=H(q,p)$
is also a constant of motion. This equality indeed describes the classical dynamical
trajectory. Therefore, in addition, this leads to another conserved quantity:
the combination of moments given by $H_Q-H(q,p)$ (and $H_C-H(q,p)$
for the classical moments).
\item In this case all equations are independent of $\hbar$.
Therefore the evolution equations for the quantum and classical moments
coincide. Hence, given the same initial data for classical and quantum probability
distributions, the value of the moments and the coordinates of the centroid will
coincide for all times.
\item Finally, as will be made explicit in Subsec. \ref{sec_evolutionuncertainty},
the combination of moments that appear in the Heisenberg inequality is conserved through evolution:
$\frac{d}{dt}[G^{2,0}G^{0,2}-(G^{1,1})^2]=0$. Therefore, if we choose a state
that saturates the uncertainty relation as initial state, it will be kept saturated during the whole evolution.
\end{itemize}

All these features explain, in this context, the possibility of constructing
exact dynamical coherent states for Hamiltonians of this form, whose centroid
follows exactly a classical (point) trajectory in phase space \cite{coherent}.

Nevertheless, a quantum harmonic oscillator is very different from a classical oscillator.
Even if, given the same initial data, the dynamical (distributional) states are similar
due to all the properties explained above, the moments corresponding to stationary
states differ as an even-power series in $\hbar$ due to the recursive relation (\ref{recursive}).
It is straightforward to see that it is necessary to make use of this recursive relation since
the algebraic equations that are obtained by equaling to zero the right-hand side
of Eq. (\ref{eq_harmonic}) are not linearly independent \cite{forthcoming}.

\subsection{Linear Hamiltonians}\label{sec_semilinear}

Another set of Hamiltonians that turn out to be very interesting due to the
properties of the equations it generates is given by those Hamiltonians
that are linear in one of the basic variables. For definiteness, let us assume
a Hamiltonian that is linear in the position $q$ but have a general dependence
on the moment $p$. Its most general form is given by,
\begin{equation}
H= q \,\varphi ( p )+ \xi ( p ),
\end{equation}
for arbitrary functions $\varphi$ and $\xi$. Note that the case given by a linear
function $\varphi$ and a quadratic function $\xi$ is included in the previous
subsection about harmonic Hamiltonians. The generalization for a Hamiltonian
linear in $p$ is straightforward.

The effective quantum Hamiltonian is given by,
\begin{eqnarray}\label{linearH}
H_Q&=& q \,\varphi ( p )+ \xi ( p )\nonumber\\ &+&\sum_{n=1}^\infty\frac{1}{(n+1)!}[q \,\varphi^{(n+1)}( p )+ \xi^{(n+1)}( p )]G^{n+1,0}\nonumber\\&+& \frac{\varphi^{(n)}( p )}{n!}G^{n,1}.
\end{eqnarray}

From here it is straightforward to obtain the equations of motion for expectation values:
\begin{eqnarray}\label{linearq}
\frac{dq}{dt}&=&q \,\varphi '( p )+ \xi' ( p )
\nonumber\\&+&\sum_{n=1}^\infty\frac{1}{(n+1)!}[q \,\varphi^{(n+2)}( p )+ \xi^{(n+2)}( p )]G^{n+1,0}\nonumber\\&+& \frac{\varphi^{(n+1)}( p )}{n!}G^{n,1},\\\label{linearp}
\frac{dp}{dt}&=&-\varphi ( p )-\sum_{n=2}^\infty\frac{1}{n!}\varphi^{(n)}( p )G^{n,0}.
\end{eqnarray}
Contrary to the harmonic Hamiltonians, in this case the equations of
motion for expectation values do get contributions from certain moments and, therefore,
the centroid of the system will not generically follow a classical orbit on the phase space.
In spite of this, the system under consideration have very peculiar properties. Let us
compute the equations of motion for the moments that appear in the equations above,
namely $G^{a,0}$ and $G^{b,1}$:
\begin{widetext}
\begin{eqnarray}\label{linearG1}
\frac{dG^{a,0}}{dt} &=& a \sum_{n=1}^\infty\frac{\varphi^{(n)}( p )}{n!}\,[G^{a-1,0}G^{n,0}
-G^{a+n-1,0}],\\\label{linearG2}
\frac{dG^{b,1}}{dt} \!&=&\sum_{n=1}^\infty\frac{1}{n!}[q \,\varphi^{(n+1)}( p )+ \xi^{(n+1)}( p )]
\left(G^{b+n,0}-G^{b,0}G^{n,0}\right)\nonumber\\
&+& \frac{\varphi^{(n)}( p )}{n!}\left[bG^{b-1,1}G^{n,0}-nG^{b,0}G^{n-1,1}+(n-b) G^{b+n-1,1}\right].
\end{eqnarray}
\end{widetext}
In obtaining these equations it is useful to note that Poisson brackets between moments
$G^{a,0}$ and $G^{b,0}$ is vanishing for any value of $a$ and $b$.
For generic functions $\varphi$ and $\xi$, these equations are quite involved. Nevertheless,
the equations for the expectation values (\ref{linearq}-\ref{linearp}), in combination with those for moments
of the form $G^{a,0}$ and $G^{b,1}$ (\ref{linearG1}-\ref{linearG2}) constitute a closed, though infinite,
system of differential equations. Note that even if the derivatives of the functions
$\varphi^{(n)}$ and $\xi^{(n)}$ are vanishing for all $n$ greater or equal to certain
value $n_{\rm max}$ the mentioned subsystem will only close by considering infinite moments.
Therefore there is no decoupled finite system.
Let us explain this in more detail. If $\varphi^{(n)}=0$ for all $n>n_{\rm max}$
only moments $G^{a,0}$ and $G^{b,1}$ up to order $n_{\rm max}$
would appear in the equations for the expectation values.
But, for instance, in the equation for $G^{n_{\rm max},0}$ higher-order moments
will appear due to the last term in Eq. (\ref{linearG1}) and thus the
finite system will not close.

In any case, this is
an interesting observation since one does not need to consider all infinite moments
in order to obtain the trajectory of the centroid of the distribution.
In fact a decoupled system is formed for expectation values ($q$, $p$), and
moments $(G^{a,0},G^{a,1},G^{a,2},\dots,G^{a,m})$ for all $a$ and a fixed $m\geq 1$.
This gives a way to solve the infinite system. One can first solve for the
system given by $(q,p,G^{a,0},G^{a,1})$. Once this is known, it is possible
to solve for all $G^{a,2}$ making use of the previous solution, and so on.
Even though, in practice this
method is hardly applicable since all those are, in principle, infinite systems. 

In addition to the commented features, there is no $\hbar$ term present in
any of the Eqs. (\ref{linearq}-\ref{linearG2}).
Therefore, the moments $C^{a,0}$ and $C^{b,1}$ as well as the expectation
values $q$ and $p$ corresponding to the classical distribution will follow the
same equations as their quantum counterparts. This means that for Hamiltonians
of the form (\ref{linearH}), the centroid of the quantum distribution will not follow a
classical orbit in the phase space but will indeed follow the trajectory given
by the centroid of the classical distribution with same initial conditions.
Thus, in these quantum systems the departure from a classical orbit on phase space
is completely due to distributional effects that are also present in the evolution
of a classical distribution.

Nevertheless, the evolution of moments that are not of the form
$G^{a,0}$ or $G^{a,1}$ will be different for the classical and quantum
distributions due to the $\hbar$ terms that appear in the quantum equations of motion.
Therefore, in order to measure pure-quantum effects it is necessary to consider the evolution
of such moments. [Note that, even if in the particular case of the equation of motion for $G^{0,2}$
there is no explicit $\hbar$ term, its evolution differs from its background counterpart $C^{0,2}$
because it is coupled with other moments that indeed present pure-quantum terms
in their equations of motion.]

Technically, the form of these equations and particularly the decoupling of
equations for the variables $(q, p, G^{a,0}, G^{b,1})$, for all $a$ and $b$, from the rest of the moments
can be traced back to the fact that the mentioned moments form a closed
Poisson algebra:
\begin{eqnarray}
\{G^{a,0},G^{b,0}\}&=&0,\\
\{G^{a,0},G^{b,1}\}&=&a (G^{a-1,0}G^{b,0}-G^{a+b-1,0}),\\
\{G^{a,1},G^{b,1}\}&=&a G^{a-1,1}G^{b,0}-b G^{a,0} G^{b-1,1}\nonumber\\&+&(b-a)G^{a+b-1,1}.
\end{eqnarray}
This does not happen if one considers also moments with a higher index in the position variable.
Already including moments of the form $G^{a,2}$ would make
the Poisson algebra not to close. It is easy to see from the general form of the
brackets (\ref{GGbrackets}) that even if the Poisson brackets between
moments of the form $G^{a,0}$ and $G^{a,1}$ with those of the form $G^{b,2}$ do close,
the brackets between two moments of the form $G^{a,2}$ give other kind of contributions. For instance:
\begin{equation}
\{G^{1,2},G^{0,2}\}=-2 \,G^{0,3}.
\end{equation}
Therefore, instead of being linear if the Hamiltonian was quadratic in
the position $q$, moments of the form $G^{a,2}$
would appear in the effective Hamiltonian. And this fact would not permit
a decoupling of the equations of motion similar to the one explained above.

Finally, let us briefly explain the particular case of a Hamiltonian that depends
only on one of the two basic variables. Following with the notation
above, this corresponds to $\varphi( p )=0$.  In this case, the procedure explained
above is very useful. The momentum $p$ and all its pure fluctuations $G^{a,0}$
turn out to be constants of motion. From Eq. (\ref{linearG2}), it is easy to solve for moments $G^{a,1}$,
obtaining a linear dependence with time,
\begin{equation}
G^{a,1}(t)=G^{a,1}(t_0)+\Phi(p,G^{a,0}) (t-t_0),
\end{equation}
with a function $\Phi$ that depends on the constants of motion. With this solution at hand
one can solve for $G^{a,2}$, which will give a dependence quadratic in time.
The procedure can be iterated for higher values of the index of the position and obtain
that a generic moment $G^{a,b}$ will have a polynomial dependence on time
of order $(t-t_0)^b$. Explicit $\hbar$ terms, that will encode the differences between
the classical and quantum evolution, first appear for moments of the form $G^{a,3}$.

\section{Inequalities for statistical moments}\label{sec_inequalities}

Any set of numbers $M^{a,b}$ does not need to correspond to statistical moments
$G^{a,b}$ of a given probability distribution function. Indeed, they need to obey
certain inequalities. In this section we will derive several set of inequalities
starting from the Cauchy-Schwarz inequality. In particular, we will
derive the generalized inequality relations for quantum moments,
that do not allow all moments to be vanishing.

This analysis has theoretical importance by its own in order
to know what kind of distributions are allowed. Furthermore, it will provide us
with additional information that constraint the values of high-order moments.
This information could be used for different purposes in many scenarios.
For example to constraint the moments corresponding to a stationary state
\cite{forthcoming}, in the construction of effective group coherent states \cite{BoTs14},
or to control the validity of a prospective numerical implementation of the system.

In addition, they could also be
used to generate physically viable random initial data in order to check
the evolution of the complete parameter space. Nonetheless, as will be
commented below, it is not completely guaranteed that if a set of numbers
$M^{a,b}$ obey the inequalities that will be derived here, they will correspond
to the moments of valid probability distribution.

By construction, we have that both for quantum and classical variables,
the moments with both even indices must be positive definite:
\begin{eqnarray}
C^{2n, 2m}&\geq &0,\\\,\, G^{2n, 2m}&\geq &0 \,\,, {\rm for}\,\, n, m\in{\mathbb N}.
\end{eqnarray}

\subsection{Inequalities for the classical moments}

In order to derive the rest of the inequalities, we will make use of the Cauchy-Schwarz inequality
that, for the classical expectation value operation, takes the following form,
\begin{equation}
\langle f\, g\rangle_c^2\leq \langle f^2 \rangle_c\,\langle g^2\rangle_c,
\end{equation}
$f$ and $g$ being any function of the position $q$ and momentum $p$.
We will make two different choices for those functions: {\emph i}/$f=(p-\langle p \rangle)^a(q-\langle q \rangle)^b$ and $g=(p-\langle p \rangle)^c(q-\langle q \rangle)^d$,
and \emph{ii}/ $f=[(q-\langle q \rangle)+(p-\langle p \rangle)]^a$ and $g=[(q-\langle q \rangle)+(p-\langle p \rangle)]^b$.
These forms of functions are chosen because, as will be
shown below, they lead to a sufficiently large class of inequalities that heavily constraint the values
of moments at different orders and give us some information about the kind of distributions
that are allowed. Nevertheless, Cauchy-Schwarz inequality is obeyed for
any pair of functions, $ f$ and $ g$, and thus we do not have the certainty that
we are saturating the complete information contained in that inequality by choosing the
functions of the forms mentioned above. That is, in principle choosing moments that obey the inequalities
we will derive will not completely guarantee that they correspond to a valid probability distribution.

The first choice, $f=(p-\langle p \rangle)^a(q-\langle q \rangle)^b$ and $g=(p-\langle p \rangle)^c(q-\langle q \rangle)^d$,
leads to the following relation:
\begin{equation}\label{ineq1}
(C^{a+c,b+d})^2\leq C^{2a,2b}C^{2c,2d},
\end{equation}
for all non-negative integers $a$, $b$, $c$, $d$. This relation can also be rewritten as, 
\begin{equation}\label{ineq1bis}
(C^{n,m})^2\leq C^{2(n-c),2(m-d)}C^{2c,2d},
\end{equation}
for all $n\leq c$ and $m\leq d$.
This is an infinite set of inequalities that, for its lowest order, reproduces
the well-known relation between the covariance $C^{1,1}:=\langle(q-\langle q\rangle)(p-\langle p\rangle) \rangle$
between two random variables, in this case $q$ and $p$,
and their corresponding variances $C^{0,2}=\langle(q-\langle q\rangle)$
and $C^{2,0}=\langle(p-\langle p\rangle)$:
\begin{equation}\label{classuncertain}
(C^{1,1})^2\leq C^{2,0}C^{0,2}.
\end{equation}
For quantum moments this particular relation, as will be explained below,
will take the form of the Heisenberg uncertainty relation.

Note that even if we choose the indices of the moment that appear on
the left-hand side of inequality (\ref{ineq1}), we are
only imposing a fix value for the sums $a+c=n$ and $b+d=m$. Once this is fixed,
the right-hand side can take $\lceil(n+1)(m+1)/2\rceil$ different forms.
Therefore, from Eq. (\ref{ineq1}) we obtain, for each $C^{n,m}$,
$\lceil(n+1)(m+1)/2\rceil$ different quadratic combinations of moments
that must be greater or equal to its square. For instance, let us choose
the moment $a+c=1$ and $b+d=2$. This particular choice provides us with the following
three independent relations:
\begin{eqnarray}
(C^{1,2})^2\leq C^{2,4},\nonumber\\
(C^{1,2})^2\leq C^{0,2}C^{2,2},\nonumber\\
(C^{1,2})^2\leq C^{0,4}C^{2,0},\label{3relations}
\end{eqnarray}
where we have made use of the fact that $C^{0,0}=1$.

The second choice, $f=[(q-\langle q \rangle)+(p-\langle p \rangle)]^a$ and $g=[(q-\langle q \rangle)+(p-\langle p \rangle)]^b$,
gives rise to the general inequality,
\begin{eqnarray}\label{ineq2}
\left[\sum_{n=0}^{a+b}\left(\!\!\begin{array}{c}
a+b\\n
\end{array}\!\!\right)
C^{n,a+b-n}\right]^2&\leq&
\nonumber\\&&
\!\!\!\!\!\!\!\!\!\!\!\!\!\!\!\!\!\!\!\!\!\!\!\!\!\!\!\!\!\!\!\!\!\!\!\!\!\!\!\!\!\!\!\!\!\!\!\!\!\!
\leq\sum_{j=0}^{2a}\sum_{k=0}^{2b}
\left(\!\!\begin{array}{c}
2a\\j
\end{array}\!\!\right)
\left(\!\!\begin{array}{c}
2b\\k
\end{array}\!\!\right)
C^{j,2a-j} C^{k,2b-k}.
\end{eqnarray}
This inequality relates the square of a given sum that contains all moments
of order ${\cal O}(a+b)$ to quadratic combinations of moments of order
${\cal O}(2a)$ and ${\cal O}(2b)$. Note that for the case $a=b$, both sides
give the same result and hence this relation trivially reduces to an equality.
In general, this inequality is independent of
the previous one (\ref{ineq1}), but in some simple cases, it is contained in there.
For instance, imposing $a=1$ and $b=0$, the left-hand side vanishes and we obtain
\begin{equation}
0\leq C^{0,2}+ 2 C^{1,1} + C^{2,0},
\end{equation}
which is trivially obeyed if these objets fulfilled relation (\ref{classuncertain}).

In summary, the most important results of this section are the two infinite collections
of inequalities, (\ref{ineq1}) and (\ref{ineq2}), that must be fulfilled by the classical
statistical moments. These relations are direct consequence of the well-known
Cauchy-Schwarz inequality but, up to the best of our knowledge,
this is the first time in the literature that they are presented in this form.

\subsection{Inequalities for quantum moments}

Previous analysis should also be reproduced for quantum moments.
In particular, we should obtain relations so that when $\hbar$ is taken to
be vanishing, they reduce to the ones obtained in the previous subsection.
As we will see, the $\hbar$ terms present in these quantum inequalities
come from applying the Weyl symmetrization to reproduce the
definition of the moments.

In this case Cauchy-Schwarz inequality is written as
\begin{equation}\label{cauchy_schwarz}
|\langle \hat f^\dagger \hat g \rangle |^2\leq\langle\hat f^\dagger \hat f \rangle \langle \hat g^\dagger \hat g \rangle,
\end{equation}
where $\hat f$ and $\hat g$ are combinations of $\hat q$ and $\hat p$ operators.
In order to generalize the inequalities derived for the classical moments,
we will consider again the two different forms of operators $\hat f$ and $\hat g$: on the one hand
$i/$ $[(\hat q-\langle \hat q \rangle)+(\hat p-\langle \hat p \rangle)]^a$ and, on the other hand,
$ii/$ $(\hat p - p)^a(\hat q-q)^b$ .

Interestingly, the first choice for the operators,
$\hat f=[(\hat q-\langle \hat q \rangle)+(\hat p-\langle \hat p \rangle)]^a$ and
$\hat g=[(\hat q-\langle \hat q \rangle)+(\hat p-\langle \hat p \rangle)]^b$, leads formally
to the same inequalities as their classical counterparts:
\begin{eqnarray}
\left[\sum_{n=0}^{a+b}\left(\!\!\begin{array}{c}
a+b\\n
\end{array}\!\!\right)
G^{n,a+b-n}\right]^2&\leq&\nonumber\\
&&\!\!\!\!\!\!\!\!\!\!\!\!\!\!\!\!\!\!\!\!\!\!\!\!\!\!\!\!\!\!\!\!\!\!\!\!\!\!\!\!\!\!\!\!\!\!\!\!\!\!
\leq\sum_{j=0}^{2a}\sum_{k=0}^{2b}
\left(\!\!\begin{array}{c}
2a\\j
\end{array}\!\!\right)
\left(\!\!\begin{array}{c}
2b\\k
\end{array}\!\!\right)
G^{j,2a-j} G^{k,2b-k},
\end{eqnarray}
which provides independent inequalities for all $b>a$.
Note that in these relations there is no presence of $\hbar$. This is due to the
fact that, for the above mentioned form of the operators, $\hat f$ and $\hat g$ are self-adjoint
and their products $\hat f \hat g$, $\hat f \hat f$ and $\hat g \hat g$ are already
in completely symmetrical Weyl ordering. Therefore, the definition of quantum moments
is obtained without any use of the commutation relations.

On the contrary, the inequalities derived from the second choice of operators $ii/$ $\hat f=(\hat p - p)^a(\hat q-q)^b$
and $\hat g=(\hat p - p)^c(\hat q-q)^d$, for different values of the powers $(a,b,c,d)$ and with different orderings, will
in general have explicit $\hbar$ terms.

Some particular cases that do not involve any $\hbar$ terms are those for which
$\hat f$ and $\hat g$ commute, for instance $\hat f=(\hat p-\langle\hat p\rangle)^a$
and $\hat g=(\hat p-\langle\hat p\rangle)^d$. In this way, we obtain the following general
relations for quantum moments, that do not involve $\hbar$ and are formally the
same as the classical inequalities:
\begin{eqnarray}\label{qineq1}
(G^{a+b,0})^2&\leq &G^{2a,0} G^{2d,0},\\\label{qineq2}
(G^{0,a+d})^2&\leq &G^{0,2a} G^{0,2b}.
\end{eqnarray}

The rest of the cases, that are not included in relations
(\ref{qineq1}-\ref{qineq2}) are more complicated due to the non-commutativity of operators
and involve new terms as a power series in $\hbar^2$. All of them can be formally written
as
\begin{eqnarray}
\left(G^{a+c,b+d}\right)^2\leq G^{2a,2b}G^{2c,2d}+\!\!\!\!
\sum_{n,i,j,k,l}\!\!\hbar^{2n}\alpha_{abcd}^{ijkl}G^{i,j}G^{k,l},
\end{eqnarray}
with certain coefficients $\alpha_{abcd}^{ijkl}$. The sum
on $n$ runs over all integers from 1 to the integer part of $(a+b+c+d)/2$ and relation
$4n+i+j+k+l=2(a+b+c+d)$ is obeyed for all terms in the sum.

It is quite difficult to obtain the analytical form of the coefficients $\alpha_{abcd}^{ijkl}$,
thus we have obtained these relations by direct computation by
making use of the iterative algebraic code presented in Ref. \cite{BBH11}.
All combinations between operators of the form $\hat f=(\hat p-\langle\hat p\rangle)^a(\hat q-\langle\hat q\rangle)^b$
and $\hat g=(\hat p-\langle\hat p\rangle)^a(\hat q-\langle\hat q\rangle)^b$,
for all internal orderings of operators $\hat q$ and $\hat p$, from first up to fifth order
(that is from $a+b=1$ up to $a+b=5$, and the same for $c+d$) have been considered.
In this way, we have obtained 1449 different inequalities that involve moments up
to tenth order. All of them might not be independent and there are several inequalities
that have the same classical limit, that is, with different coefficients $\alpha_{abcd}^{ijkl}$.
In those cases, one should take the most restrictive ones but this is not a simple analysis
with so many variables. Therefore, for practical reasons, we keep all of them.
In order to illustrate the form of these quantum inequalities here we will just show the quantum
generalization of the particular examples shown in the previous subsection. In addition, in
the next subsection, a particularly important set of inequalities will be analyzed: the uncertainty relations.

Regarding the examples shown explicitly in the previous subsection, on the one hand the classical
relation between the covariance and the variance (\ref{classuncertain}) is generalized by
the Heisenberg uncertainty principle:
\begin{equation}\label{uncert}
(G^{1,1})^2\leq  G^{2,0}G^{0,2} -\frac{\hbar^2}{4}\,.
\end{equation}

On the other hand relations (\ref{3relations}), derived from (\ref{ineq1}) for a left-hand side of
the form $(C^{1,1})^2$, are generalized by the following inequalities between quantum
moments:
\begin{eqnarray}
(G^{1,2})^2&\leq& G^{2,4} + \hbar^2 G^{0,2},\nonumber\\
(G^{1,2})^2&\leq& G^{0,2} G^{2,2} + \frac{\hbar^2}{2} G^{0,2},\nonumber\\
(G^{1,2})^2&\leq&G^{0,4} G^{2,0}.\nonumber
\end{eqnarray}
As can be seen, some inequalities, as the last one, do not change and
keep the same form as the one corresponding to the classical moments.

\subsection{Uncertainty relations}

It is interesting to note that in the classical setting, a Dirac delta distribution is
consistent with all derived inequalities. That is, it is possible to take all $C^{a,b}$ moments
to be vanishing. On the contrary, as it is well known from the Heisenberg uncertainty principle, in the quantum case
the limit $G^{a,b}\rightarrow 0$ is not consistent.

In fact, most of derived inequalities for quantum moments allow for a distribution given
by Dirac delta; that is, they are not violated if one imposes all quantum moments
to be vanishing. Heisenberg uncertainty relation (\ref{uncert}) is the most simple inequality that does not
allow for such a limit. In addition to that, among the 1449 different relations that have been obtained,
there are 160 relations that forbids the mentioned limit. These relations are the ones that,
as well as the uncertainty relation, encode the necessary
lack of information of the quantum system an observer must have.

All uncertainty relations come from the quantum version of the classical
inequalities (\ref{ineq1bis}) with $n=m$ and can be formally written as
\begin{eqnarray}\label{uncertaintygeneric}
(G^{n,n})^2+\gamma_{ncd} \hbar^{2n}&\leq& G^{2(n- c), 2(n-d)} G^{2c, 2d}+
\nonumber\\&+&\sum_{i=1}^{n-1}\alpha_{nijklm} \hbar^{2i}G^{j,k}G^{l,m}.
\end{eqnarray}
Interestingly the uncertainty relations are only those inequalities for which $c\neq d$, since in that case the
constant $\gamma_{ncd}$ turns out to be positive. The case $c=d=n/2$, for even $n$, trivially reduces to an equality;
whereas for the rest of the $c=d$ cases, $\gamma_{ncd}$ is a negative constant and thus
does not constitute an uncertainty relation since it is obeyed when choosing all $G^{a,b}$ moments to be vanishing.
Uncertainty relations with $n=1$ and $n=2$ are explicitly shown in the appendix.

Essentially, all these uncertainty relations bound from below
products of the form $G^{2a,0}G^{0,2a}$. In fact, it is possible to choose all moments vanishing
except $G^{2a,0}$ and $G^{0,2a}$. Under this particular choice, the 160 uncertainty relations reduce
to five inequalities of the form,
\begin{equation}\label{uncertaintyonly2n0}
\gamma_n\hbar^{2n}\leq G^{0,2n}G^{2n,0},
\end{equation}
with a positive constant $\gamma_n$ for each value of $n$. More precisely: $\gamma_1=1/4$,
$\gamma_2=3/8$, $\gamma_3=81/64$, $\gamma_4=9/4$, and $\gamma_5=225/16$.

Following with this choice of moments (all vanishing except $G^{2a,0}$ and $G^{0,2a}$),
one could choose to have the
same uncertainty both in momentum and position by imposing $G^{2a,0}=G^{0,2a}= g_a \hbar^a$ and
uncertainty relations (\ref{uncertaintyonly2n0}) will simply impose a minimum value for each adimensional
moment $g_n$. In addition to those inequalities,
it is also necessary to take into account the
rest of the quantum inequalities that have been derived in the previous section.
In particular, it is interesting to note that with this simple choice
all quantum inequalities that contain up to eighth-order moments
are reduced to the following constraints on the adimensional moments $g_a$:
$$
g_1\geq\frac{1}{2},\,\,\,g_2\geq6(g_1)^2,\,\,\,g_3g_1\geq\frac{9}{4}(g_2)^2,\,\,\,g_4g_2\geq\frac{25}{14}(g_3)^2.
$$
If one wishes to find a minimum value for the moments, as can be seen in this expression,
one would have to face the tension that exist between lower-order moments and higher-order ones.
For instance, from the third inequality
with a fixed value of $g_2$, the lower $g_1$ the larger $g_3$ will have to be chosen.
In any case,
one could choose the saturation of every inequality starting from the lowest one, which is not affected by higher-order
moments, and solve the inequalities order by order.
Nevertheless, as has been commented on the previous subsection, there is no certainty that
solving all derived inequalities will lead to a valid state since there could be other
constraints on the moments that have not been considered.

\subsection{Other allowed distributions}

Apart from the vanishing moments limit of a Dirac delta,
there are other distribution of moments that are allowed by the classical inequalities we have derived,
but not by the quantum ones. This is the case for instance of $C^{a,b}=\hbar^{(a+b)/2}$, or
$C^{a,b}=A^{a+b}\hbar^{(a+b)/2}$, with an adimensional constant $A$. (The factor $\hbar^{(a+b)/2}$
stands for dimensional reasons.)

Apparently, in order to fulfill the corresponding inequalities,
the quantum moments need a factor in front of $\hbar^{(a+b)/2}$ that increases faster than $A^{a+b}$ with the indices $a$ and $b$.
In particular, the following distributions are allowed both by classical and quantum
inequalities:
\begin{eqnarray}
G^{a,b}=a! b! \hbar^{(a + b)/2},\\
G^{a,b}=a^a b^b \hbar^{(a + b)/2},\\
G^{a,b}=a^{a-1} b^{b-1} \hbar^{(a + b)/2}.
\end{eqnarray}
Interestingly the distribution $G^{a,b}=a^{a - 2} b^{b - 2} \hbar^{(a + b)/2}$ is
allowed by classical but not by quantum inequalities; whereas $G^{a,b}=a^{a - 3} b^{b - 3} \hbar^{(a + b)/2}$
is neither allowed by quantum nor by classical relations.

On the other hand, defining the moments as functions of their order only, like $G^{a,b}=(a + b)! \hbar^{(a + b)/2}$
obey all inequalities except Heisenberg uncertainty relation.

Finally all these considerations leads to think that the quantum moments necessarily form a divergent series.
If we assume that this divergent series is asymptotic, it is possible to analyze up to which order
they converge, that is, which will the smallest term of the series. Usually the truncation at that
order is the optimal truncation in the sense that the accuracy of the result improves as one includes
more and more orders up to the mentioned smallest term. From that order on, adding more orders will
worsen the result. For instance, for the distribution given by
$a! b! \hbar^{(a + b)/2}$, two moments of consecutive orders
$G^{n-1, 0}$ and $G^{n, 0}$ will be of the same order when $n=1/\sqrt{\hbar}$.
Therefore, the smaller the value of $\hbar$ is chosen, the more orders can be considered
in order to improve the result.

\subsection{Evolution of the uncertainty relations}\label{sec_evolutionuncertainty}

Let us analyze the evolution of the Heisenberg uncertainty principle. Given a generic Hamiltonian $H(q,p)$,
it is straightforward to obtain
\begin{widetext}
\begin{eqnarray}\label{evHeisenberg}
\frac{\md}{\md t} [G^{2,0}G^{0,2}-(G^{1,1})^2]=
\sum_{a+b\geq3}\frac{2}{a!b!}\frac{\partial^{a+b} H}{\partial p^a\partial q^b}[a\, G^{2,0} G^{a-1,b+1}-
b\, G^{0,2}G^{a+1,b-1}+(b-a)\,G^{1,1}G^{a,b}],
\end{eqnarray}
\end{widetext}
where the sum is for all non-negative $a$ and $b$ that obeys the constraint $a+b\geq3$.
It is interesting to note
that in the right-hand side of that expression only derivatives of the Hamiltonian higher
than two appear. In the evolution equation for each of the moments second-order derivatives
of the Hamiltonian are present but, for that specific combination of moments, these terms cancel each other.
Therefore, this leads to another important property of harmonic Hamiltonians:
the combination of moments in the Heisenberg uncertainty principle is a constant of motion.
Certainly, this does not mean that the initial state will not be deformed through evolution.
All moments will generically grow in absolute value with time but keeping that relation
constant. For squeezed states, defined as those that saturate Heisenberg uncertainty relation,
this saturation will be kept throughout evolution. However, this property does not apply
to the higher-order uncertainty relations that have been presented in the previous subsection.
The combination of moments appearing in those relations are not generically conserved even
for harmonic Hamiltonians.

On the other hand, as a general feature of second-order moment evolution,
there are no $\hbar$ terms present in Eq. (\ref{evHeisenberg}). Thus for second-order moments the classical and quantum equations of motion are exactly the same, and
the above considerations also apply to the relation $C^{0,2}C^{2,0}-(C^{1,1})^2\geq0$ between classical moments.
This, of course, does not mean that considering the full dynamics under a generic non-harmonic Hamiltonian
the behavior of the classical and quantum system will be the same. The quantum corrections enter at third order
and, due to the back-reaction, they will also affect the evolution of background quantities and second-order moments.

\section{Conclusions}\label{sec_conclusions}

Quantum effects have two different origins. On the one hand, \emph{distributional effects} are a direct
consequence of the Heisenberg uncertainty relation, since the state of a quantum system
is necessarily given by an extended probability distribution, instead of a single point in phase space.
Technically this means that the statistical moments that describe this distribution can not be
vanishing, and they back-react on the evolution of the position of the centroid.
Thus generically the centroid of a quantum distribution does not follow a classical (point)
trajectory in the phase space. On the other hand, due to the non-commutativity of the basic
quantum operators, the \emph{non-commutative} effects appear
in the evolution equations of quantum statistical moments as terms
that come with an explicit dependence on $\hbar$. Due to the smallness of this constant, the distributional effects may
be more important than these latter ones.
In the formalism used in this paper, based on a decomposition on statistical moments of the
wave function, the distinction between these two
sources of quantum corrections is very neat.

Furthermore, the evolution of a classical probability distribution
has also been considered in terms of its statistical moments. This is particularly important
since, even if in the context of classical mechanics is ideally possible to study the evolution
of an initial point in the phase space, in practice there are always errors in the initial conditions
and thus one needs to deal with an extended probability distribution. In the classical case,
the distributional effects are also present and generically the centroid of a distribution does not
either follow a classical orbit in the phase space. In fact, the only difference between the quantum
and classical evolution equations are the explicit $\hbar$ factors that appear in the quantum
system. Therefore, when taking the $\hbar\longrightarrow0$ limit on the quantum system,
one does not recover a unique classical trajectory on the phase space, but an ensemble of them.

There are some Hamiltonians that have very peculiar properties regarding the
classical and quantum evolution they generate. The harmonic Hamiltonians, those
that are at most quadratic on the basic variables, are very special. One of their most
important property is that they generate the same dynamics
both in the classical and quantum sector. The Hamiltonians that are linear in one
of the basic variables also generate the same classical and quantum flux
for the position of the centroid of the distribution and certain infinite set of moments. Therefore, in order to look for pure-quantum
effects in this kind of systems, one should check the evolution of moments that are
not contained in that set.

Finally, a large set of inequalities obeyed by classical and quantum moments have
been obtained by making use of the Cauchy-Schwarz inequality. Among these,
the uncertainty relations obeyed by quantum moments, that are defined as those
that do not allow for all moments to be vanishing. The simplest of these relations
is the well-known Heisenberg uncertainty principle. The rest constitute its higher-order
generalizations. In essence they
provide additional information that constraint the value of high-order moments.
This information could be used in different analysis;
for instance, to constraint the value of the moments corresponding to a stationary state
\cite{forthcoming}, in the construction of effective group coherent states \cite{BoTs14},
or to control the validity of a prospective numerical implementation of the system.
The distribution of moments allowed by these inequalities have also been analyzed,
as well as the evolution of the uncertainty relations under generic Hamiltonians.

As future work, the formalism developed in this article will be applied to
different simple physical systems, like the harmonic and quartic oscillators, in order
to measure the strength of each kind of quantum effect and to obtain the statistical
moments corresponding to their (classical and quantum) stationary states \cite{forthcoming}. It would also
be interesting to revisit the different cosmological models commented in the introduction,
like the ones studied in \cite{BoTa08, BBH11, Boj07}, to compare their quantum and classical distributional evolution.

\acknowledgements

The author thanks Carlos Barcel\'o, Ra\'ul Carballo-Rubio,
I\~naki Garay, Claus Kiefer, Manuel Kr\"amer, and Hannes Schenck
for discussions and comments. Special thanks to Martin Bojowald for interesting comments
on a previous version of this manuscript. Financial support from the Alexander von
Humboldt Foundation through a postdoctoral fellowship is gratefully acknowledged.
This work is supported in part by Projects IT592-13 of the Basque Government
and FIS2012-34379 of the Spanish Ministry of Economy and Competitiveness.

\appendix

\section{Uncertainty relations}

In order not to extend
this appendix in excess, here the uncertainty relations that contain
moments only up to sixth order are presented.
The generalized uncertainty relations, defined as those inequalities that are not obeyed 
for the case $G^{a,b}=0$, are the quantum equivalent of the inequalities (\ref{ineq1}) for
classical moments and can be generically written as shown in (\ref{uncertaintygeneric}).
In particular, the well-known Heisenberg uncertainty relation
is given as,
\begin{equation}
 \hbar^2/4 + (G^{1, 1})^2 \leq G^{0, 2} G^{2, 0}.
\end{equation}

The generalizations of $(C^{2,2})^2\leq C^{2a,2c} C^{2b,2d}$ with $a+b=2$ and $c+d=2$ gives rise to
the following inequalities:
\begin{eqnarray*}
 \hbar^4/4 + (G^{2, 2})^2 &\leq & G^{2, 0} G^{2, 4}+\hbar^2 [3 G^{0, 2} G^{2, 0} - G^{2, 2}], 
 \\ \hbar^4/4 + (G^{2, 2})^2  &\leq  &  G^{2, 0} G^{2, 4}+
\\\nonumber &+& \hbar^2 [G^{0, 2} G^{2, 0} + G^{2, 2}-4 (G^{1, 1})^2], 
\\ \hbar^4/4  + (G^{2, 2})^2 &\leq & G^{0, 4} G^{4, 0}+ \hbar^2 [G^{2, 2}-4 (G^{1, 1})^2], 
 \\ \hbar^4/4  + (G^{2, 2})^2 &\leq & G^{0, 2} G^{4, 2} + \hbar^2 [3 G^{0, 2} G^{2, 0} - G^{2, 2}], 
\\  \hbar^4/4 + (G^{2, 2})^2 &\leq & G^{0, 2} G^{4, 2}+
\\&+& \hbar^2 [G^{0, 2} G^{2, 0} + G^{2, 2}-4 (G^{1, 1})^2] .
\end{eqnarray*} 
Note that the first two relations (as well as the last two) are identical except for the coefficient of $\hbar^2$.
Summing them up, it is possible to define the relation
$$\hbar^4/4  + (G^{2, 2})^2 \leq G^{2,0} G^{2,4}+ 2\hbar^2 [G^{2,0} G^{0, 2}-(G^{1, 1})^2],$$
where the combination of moments inside the square bracket is the same that appears in the Heisenberg
uncertainty principle. The same can be done with the last couple of inequalities.
Note that, as explained in Sec. IVC, the fourth classical relation derived from that inequality,
$(C^{2,2})^2\leq C^{4,4}$, does not give rise to a quantum uncertainty relation because
the corresponding $\gamma_{ncd}$ coefficient (\ref{uncertaintygeneric}) is negative and thus
allows the $G^{a,b}\rightarrow 0$ limit.

The uncertainty relations at next order are the quantum analog of the classical relations
(\ref{ineq1}) with $n=m=3$:
\begin{eqnarray*}
(C^{3,3})^2\leq C^{4,2} C^{2,4}, &\qquad &(C^{3,3})^2\leq C^{6,0} C^{0,6},\\
(C^{3,3})^2\leq C^{4,0}C^{2,6} , &\qquad &(C^{3,3})^2\leq C^{6,2} C^{0,4},\\
(C^{3,3})^2\leq C^{4,4} C^{2,2}, &\qquad &(C^{3,3})^2\leq C^{6,4} C^{0,2},\\
(C^{3,3})^2\leq C^{4,6} C^{2,0}, &\qquad& (C^{3,3})^2\leq C^{6,6} .\\
\end{eqnarray*}
All these relations are converted into quantum uncertainty relations except for
$(C^{3,3})^2 \leq C^{4,4}C^{2,2}$ and $(C^{3,3})^2\leq C^{6,6}$ that
has a negative $\gamma_{ncd}$ coefficient (\ref{uncertaintygeneric}) in the
free term. Nonetheless, only the quantum generalization
of the first two relations will be explicitly shown since the rest involve moments of an order higher than six.
These are given as follows:
\begin{widetext}
\begin{eqnarray*} 
9 \hbar^6/16 + (G^{3, 3})^2&\leq&    G^{2, 4} G^{4, 2}
+   \hbar^4 [3 G^{0, 2} G^{2, 0} - 9/4 ((G^{1, 1})^2 + G^{2, 2})] + 
\\  &+&
\hbar^2 [-9/4 (G^{2, 2})^2 + 3 G^{2, 0} G^{2, 4} - 3 G^{1, 1} G^{3, 3} + 
      G^{0, 2} G^{4, 2}],
 \\ \hbar^6/16  + (G^{3, 3})^2 &\leq& G^{2, 4} G^{4, 2} +
   \hbar^4 [-9/4 (G^{1, 1})^2 + 3 G^{0, 2} G^{2, 0} - 1/4 G^{2, 2}] + 
\\ &+&   \hbar^2 [-(G^{2, 2})^2/4 + 3 G^{2, 0} G^{2, 4} - 3 G^{1, 1} G^{3, 3} + 
      G^{0, 2} G^{4, 2}], 
 \\  \hbar^6/16 + (G^{3, 3})^2 &\leq&  G^{2, 4} G^{4, 2}+
   \hbar^4 [- (G^{1, 1})^2/4 + G^{0, 2} G^{2, 0} - 3/4 G^{2, 2}]  + 
\\&+&   \hbar^2 [-9/4 (G^{2, 2})^2 + G^{2, 0} G^{2, 4} - G^{1, 1} G^{3, 3} + 
      G^{0, 2} G^{4, 2}], 
 \\ \hbar^6/16 + (G^{3, 3})^2 &\leq&  G^{2, 4} G^{4, 2}+
   \hbar^4 [-(G^{1, 1})^2/4 + G^{0, 2} G^{2, 0} - 5/4 G^{2, 2}]  + 
\\&+&   \hbar^2 [-25/4 (G^{2, 2})^2 + G^{2, 0} G^{2, 4} + G^{1, 1} G^{3, 3} + 
      G^{0, 2} G^{4, 2}], 
 \\ \hbar^6/16 + 
   (G^{3, 3})^2 &\leq& G^{2, 4} G^{4, 2}+
   \hbar^4 [-25/4 (G^{1, 1})^2 + G^{0, 2} G^{2, 0} + 7/4 G^{2, 2}]  + 
\\&+&   \hbar^2 [-49/4 (G^{2, 2})^2 + G^{2, 0} G^{2, 4} + 5 G^{1, 1} G^{3, 3} + 
      G^{0, 2} G^{4, 2}], 
 \\  9 \hbar^6/16 + 
   (G^{3, 3})^2 &\leq& G^{2, 4} G^{4, 2} +
   \hbar^4 [3 G^{0, 2} G^{2, 0} - 9/4 ((G^{1, 1})^2 + G^{2, 2})]
   \\&+& \hbar^2 [-9/4 (G^{2, 2})^2 + G^{2, 0} G^{2, 4} - 3 G^{1, 1} G^{3, 3} + 
      3 G^{0, 2} G^{4, 2}], 
 \\  \hbar^6/16+ (G^{3, 3})^2 &\leq&  G^{2, 4} G^{4, 2}+
   \hbar^4 [-9/4 (G^{1, 1})^2 + 3 G^{0, 2} G^{2, 0} - 1/4 G^{2, 2}]   + 
\\&+&   \hbar^2 [-(G^{2, 2})^2/4 + G^{2, 0} G^{2, 4} - 3 G^{1, 1} G^{3, 3} + 
      3 G^{0, 2} G^{4, 2}], 
 \\ \hbar^6/16 + (G^{3, 3})^2 &\leq& G^{2, 4} G^{4, 2} +
   \hbar^4 [-9/4 (G^{1, 1})^2 + 9 G^{0, 2} G^{2, 0} - 1/4 G^{2, 2}] + 
\\&+&   \hbar^2 [-(G^{2, 2})^2/4 + 3 G^{2, 0} G^{2, 4} - 3 G^{1, 1} G^{3, 3} + 
      3 G^{0, 2} G^{4, 2}], 
 \\ 9 \hbar^6/16+ (G^{3, 3})^2  &\leq& 
   G^{0, 6} G^{6, 0} - 27/4 \hbar^4 [3 (G^{1, 1})^2 - G^{2, 2}] + \hbar^2 [-81/4 (G^{2, 2})^2 + 9 G^{1, 1} G^{3, 3}] .
\end{eqnarray*}
\end{widetext}


\begin{thebibliography}{100}

\bibitem{BoTr88}
W. Boucher and J. H . Traschen, Phys. Rev. D {\bf 37}, 3522 (1988).

\bibitem{And95}
A. Anderson, Phys. Rev. Lett. {\bf 74}, 621 (1995).

\bibitem{ShSu78}
T. N. Sherry and E. C. G. Sudarshan, Phys. Rev. D {\bf 18}, 4580 (1978); Phys. Rev. D {\bf 20}, 857 (1979).

\bibitem{PeTe01}
A. Peres and D. R. Terno, Phys. Rev. A {\bf 63}, 022101 (2001).

\bibitem{Elz12}
H. T. Elze, Phys. Rev. A {\bf 85}, 052109 (2012).

\bibitem{CHS12}
A. J. K. Chua, M. J. W. Hall, and C. M. Savage, Phys. Rev. A {\bf 85}, 022110 (2012).

\bibitem{CaSa99}
J. Caro and L. L. Salcedo, Phys. Rev. A {\bf 60}, 842 (1999).

\bibitem{BCG12}
C. Barcel\'o, R. Carballo-Rubio, L. J. Garay, and R. G\'omez-Escalante, Phys. Rev. A {\bf 86}, 042120 (2012).

\bibitem{BYZ94}
L. E. Ballentine, Y. Yang, and J. P. Zibin, Phys. Rev. A {\bf 50}, 2854 (1994).

\bibitem{BM98}
L. E. Ballentine and S. M. McRae, Phys. Rev. A {\bf 58}, 1799 (1998).

\bibitem{BoSk06}
M. Bojowald and A. Skirzewski, Rev.\ Math.\ Phys. {\bf 18},  713  (2006).

\bibitem{Boj12}
M. Bojowald, Class. Quantum Grav. {\bf 29}, 213001 (2012).

\bibitem{BoTs09}
M. Bojowald and A. Tsobanjan, Phys.\ Rev.\ D {\bf 80},  125008  (2009); Class.\ Quantum Grav. {\bf 27},  145004 (2010).

\bibitem{BoTa08}
M. Bojowald and R. Tavakol, Phys.\ Rev.\ D {\bf 78},  023515  (2008).

\bibitem{BBH11}
M. Bojowald, D. Brizuela, H. H. Hern\'andez, M. J. Koop, and H. A. Morales-T\'ecotl,
Phys. Rev. D {\bf 84} 043514 (2011).

\bibitem{Boj07}
M. Bojowald, Phys. Rev. D {\bf 75}, 081301 (2007); Phys. Rev. D {\bf 75} 123512 (2007).

\bibitem{BHT11}
M. Bojowald, P.~A. H\"ohn, and A. Tsobanjan, Class.\ Quantum Grav. 
{\bf 28}, 035005 (2011); Phys. Rev. D {\bf 83}, 125023 (2011).

\bibitem{HKT12}
P. A. H\"ohn, E. Kubalova, and A. Tsobanjan, Phys. Rev. D {\bf 86}, 065014 (2012).

\bibitem{BSS09}
M. Bojowald, B. Sandh\"ofer, A. Skirzewski, and A. Tsobanjan,
Rev.\ Math.\ Phys. {\bf 21},  111  (2009).

\bibitem{forthcoming}
D. Brizuela, \emph{in preparation}.

\bibitem{Ban77}
K. Banerjee, Phys. Lett. A {\bf 63}, 223 (1977).

\bibitem{RiBl79}
J. L. Richardson and R. Blankenbecler, Phys. Rev. D {\bf 19}, 496 (1979).

\bibitem{BrSu81}
J. B. Bronzan and R. L. Sugar, Phys. Rev. D {\bf 23}, 1806 (1981).

\bibitem{Has78}
R. W. Hasse, J. Phys. A: Math. Gen. {\bf 11}, 1245 (1978). 

\bibitem{ReHe80}
B. Remaud and E. S. Hern\'andez, J. Phys. A: Math. Gen. {\bf 13} 2013 (1980). 

\bibitem{AnHa85}
M. Andrews and M. Hall, J. Phys. A: Math. Gen. {\bf 18}, 37 (1985).

\bibitem{coherent}
R. Glauber, Phys. Rev. {\bf 131}, 2766 (1963).

\bibitem{BoTs14}
M. Bojowald and A. Tsobanjan, Class. Quant. Grav. {\bf 31}, 115006 (2014).

\end{thebibliography}
\end{document}